%% file: SMP-22-012_temp.tex
\pdfoutput=1
\documentclass[11pt,twoside,a4paper,cmspaper,final,collab]{cms-tdr}
\def\svnVersion{8f8bd73}\def\svnDate{2025/04/11}\def\cmsCernNoTag{CERN-EP-2024-253}\def\cmsCernDate{\today}\def\cmsMessage{Published in Physics Letters B as \href{http://dx.doi.org/10.1016/j.physletb.2025.139462}{\doi{10.1016/j.physletb.2025.139462}.}}
\begin{document}\cmsNoteHeader{SMP-22-012}

\newlength\cmsFigWidth\ifthenelse{\boolean{cms@external}}{\setlength\cmsFigWidth{0.45\textwidth}}{\setlength\cmsFigWidth{0.65\textwidth}}
\ifthenelse{\boolean{cms@external}}{\providecommand{\cmsLeft}{top\xspace}}{\providecommand{\cmsLeft}{left\xspace}}
\ifthenelse{\boolean{cms@external}}{\providecommand{\cmsRight}{bottom\xspace}}{\providecommand{\cmsRight}{right\xspace}}
\ifthenelse{\boolean{cms@external}}{\providecommand{\cmsTableMaybe}[1]{#1}}{\providecommand{\cmsTableMaybe}[1]{\resizebox{\textwidth}{!}{#1}}}
\providecommand{\cmsTable}[1]{\resizebox{\textwidth}{!}{#1}}
\newlength\cmsTabSkip\setlength{\cmsTabSkip}{1ex}
\newcommand{\mmm}{\ensuremath{m_{\mu\mu}}\xspace}
\newcommand{\mmmg}{\ensuremath{m_{\mu\mu\PGg}}\xspace}
\newcommand{\kc}{\ensuremath{\kappa_{\PQc}}\xspace}
\newcommand{\kg}{\ensuremath{\kappa_{\PGg}}\xspace}
\newcommand{\PGyPnS}{\ensuremath{\PGyP{nS}}\xspace}

\cmsNoteHeader{SMP-22-012}

\title{Search for rare decays of the \texorpdfstring{\PZ}{Z} and Higgs bosons to a \texorpdfstring{\PJGy}{J/psi} or \texorpdfstring{\PGyP{2S}}{psi(2S)} meson and a photon in proton-proton collisions at \texorpdfstring{$\sqrt{s} = 13\TeV$}{sqrt(s) = 13 TeV}}

\abstract{A search is presented for rare decays of the \PZ and Higgs bosons to a photon and a \PJGy or a \PGyP{2S} meson, with the charmonium state subsequentially decaying to a pair of muons. The data set corresponds to an integrated luminosity of 123\fbinv of proton-proton collisions at a center-of-mass energy of 13\TeV collected with the CMS detector at the LHC. No evidence for branching fractions of these rare decay channels larger than predicted in the standard model is observed. Upper limits at 95\% confidence level are set: $\mathcal{B}(\PH\to\PJGy\PGg) < 2.6 \times 10^{-4}$, $\mathcal{B}(\PH\to\PGyP{2S}\PGg) < 9.9 \times 10^{-4}$, $\mathcal{B}(\PZ\to\PJGy\PGg) < 0.6 \times 10^{-6}$, and $\mathcal{B}(\PZ\to\PGyP{2S}\PGg) < 1.3 \times 10^{-6}$. The ratio of the Higgs boson coupling modifiers $\kc/\kg$ is constrained to be in the interval $(-157,\, +199)$ at 95\% confidence level. Assuming $\kg=1$, this interval becomes $(-166,\, +208)$.}

\hypersetup{
pdfauthor={CMS Collaboration},
pdftitle={Search for the rare decays of the Z and Higgs bosons to a J/psi or psi(2S) meson and a photon in proton-proton collisions at sqrt(s)= 13 TeV},
pdfsubject={CMS},
pdfkeywords={CMS, SMP, Higgs, J/psi, photon, rare decays}
}

\maketitle

\section{Introduction}

\begin{figure}[!htb]\centering
  \includegraphics[width=0.48\cmsFigWidth]{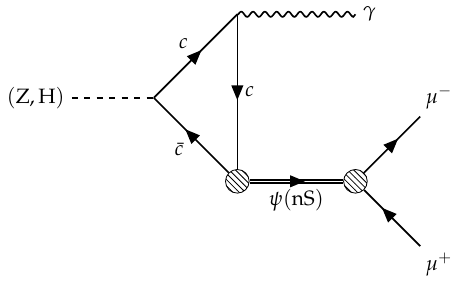}
  \includegraphics[width=0.52\cmsFigWidth]{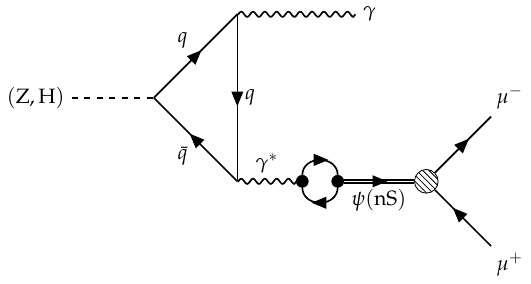} \\
  \includegraphics[width=0.50\cmsFigWidth]{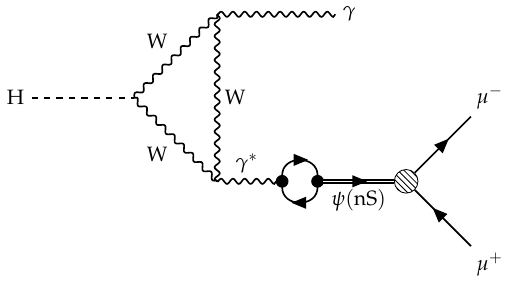}
  \includegraphics[width=0.46\cmsFigWidth]{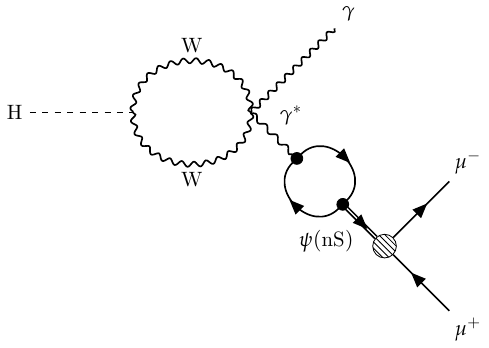}
  \caption{Leading-order Feynman diagrams of \PZ and Higgs boson rare decays to \PJGy or \PGyP{2S}, and a photon, through the direct (upper \cmsLeft) and indirect (upper \cmsRight, lower \cmsLeft, and lower \cmsRight) processes. These four diagrams are considered as the signal of this analysis.}\label{fig:signal}
\end{figure}

The discovery of the Higgs boson by the ATLAS and CMS Collaborations in 2012~\cite{bib:Aad20121,bib:Chatrchyan201230,Chatrchyan:2013lba} represented a milestone in the understanding of the electroweak sector of the standard model (SM) of elementary particles~\cite{PhysRevLett.13.321,PhysRevLett.13.508}, and it started a new era of precision measurements of the couplings of the Higgs boson to the other SM bosons and fermions~\cite{CMS:2022dwd,ATLAS:2022vkf}. New particles, postulated by theories that extend the SM, may contribute to quantum loops and modify the branching fractions of rare decays of the Higgs boson.
A channel which would be sensitive to such contributions is the decay of the Higgs boson to a photon (\PGg) and a \PJGy or \PGyP{2S} meson, collectively referred to as \PGyPnS, followed by the decay of the \PGyPnS meson to a pair of muons.
The $\PH\to\PGyPnS\PGg$ decay could take place through \cPqc quark loops, as shown in Fig.~\ref{fig:signal}, and thus an accurate measurement of this process would constrain the couplings of the Higgs boson to \cPqc quarks, providing complementary information to searches for Higgs boson decay to a pair of \PQc quarks hadronizing into jets~\cite{PhysRevLett.131.061801,PhysRevLett.131.041801,ATLAS:2022ers}. Although the latter has a better sensitivity to the coupling between the Higgs boson and the charm quark, the loop-mediated $\PH\to\PGyPnS\PGg$ is also sensitive to additional contributions that can arise in certain extensions of the SM~\cite{Bodwin:2013gca,Delaunay:2013pja}.
Instead, the yet outstanding observation of the $\PZ\to\PGyPnS\PGg$ process would represent an objective that is achievable with the integrated luminosity of the HL-LHC~\cite{dEnterria:2023wjq}.

The boson decays to \PGyPnS are considered as ``rare'', as the branching fractions range between ${\approx}10^{-6}$ and ${\approx}10^{-8}$~\cite{LHCHiggsCrossSectionWorkingGroup:2016ypw,Bodwin:2013gca,Bodwin:2014bpa,Huang:2014cxa,Konig:2015qat,Bodwin:2017pzj,Bodwin:2017wdu,Brambilla:2019fmu,Sang:2023hjl,Wang:2023ssg,dEnterria:2023wjq}.
On the other hand, the branching fractions of the decays of the \PGyPnS to a pair of muons have been measured with great precision~\cite{Navas:2024pdg}.
The values considered in this analysis are reported in Table~\ref{tab:br}.

\begin{table}[!htp]\centering
    \topcaption{The SM predictions of the branching fractions of the Higgs and \PZ boson decays to $\PGyPnS\PGg$ used in this analysis, and measured branching fractions for the $\PGyPnS\to\PGm\PGm$ decays.}
    \label{tab:br}
    \begin{tabular}{lcc}
        Process                     & $\mathcal{B}$                         & Ref. \\
        \hline
        $\PH\to\PJGy\PGg$           & $(3.01 \pm 0.15) \times 10^{-6}$      & \cite{Brambilla:2019fmu}  \\
        $\PZ\to\PJGy\PGg$           & $8.96^{+1.51}_{-1.38} \times 10^{-8}$ & \cite{Bodwin:2017pzj}     \\
        $\PH\to\PGyP{2S}\PGg$       & $(1.03 \pm 0.06) \times 10^{-6}$      & \cite{Brambilla:2019fmu}  \\
        $\PZ\to\PGyP{2S}\PGg$       & $4.83^{+1.02}_{-0.91} \times 10^{-8}$ & \cite{Bodwin:2017pzj}     \\
        [\cmsTabSkip]
        $\PJGy\to\PGmm\PGmp$        & $(5.961 \pm 0.033) \times 10^{-2}$    & \cite{Navas:2024pdg} \\
        $\PGyP{2S}\to\PGmm\PGmp$    & $(8.0 \pm 0.6) \times 10^{-3}$        & \cite{Navas:2024pdg} \\
    \end{tabular}
\end{table}

The search for the decays $\PZ\to\PGyPnS\PGg$ and $\PH\to\PGyPnS\PGg$ was previously performed by both the ATLAS and CMS Collaborations using data collected in 2016~\cite{CMS:2018gcm,ATLAS:2018xfc} and the ATLAS Collaboration recently updated its search using the complete Run 2 data set~\cite{ATLAS:2022rej}.
The present search is performed with data from proton-proton collisions at a center-of-mass energy of $13\TeV$, corresponding to an integrated luminosity of 123\fbinv, collected with the CMS detector during the LHC Run 2 (2016--2018).

Tabulated results are provided in the HEPData record for this analysis~\cite{HEPData}.

\section{The CMS detector and event reconstruction}\label{sec:detector}

A detailed description of the CMS detector, together with a definition of the coordinate system used and the relevant kinematic variables, is reported in Ref.~\cite{Chatrchyan:2008zzk}.

The central feature of the CMS apparatus is a superconducting solenoid of 6\unit{m} internal diameter, providing a magnetic field of 3.8\unit{T}. Within the magnetic volume are a silicon pixel and strip tracker, a lead tungstate crystal electromagnetic calorimeter (ECAL), and a brass and scintillator hadron calorimeter (HCAL), each composed of a barrel and two endcap sections. The ECAL and HCAL provide pseudorapidity coverage up to $\abs{\eta}<3.0$, which is further extended by forward calorimeters to $\abs{\eta}<5.0$~\cite{Chatrchyan:2008zzk}. Muons are measured in gas-ionization detectors embedded in the steel flux-return yoke outside the solenoid.

Events of interest are selected using a two-tiered trigger system. The first level, composed of custom hardware processors, uses information from the calorimeters and muon detectors to select events at a rate of around 100\unit{kHz} within a fixed latency of 4\mus~\cite{Sirunyan:2020zal}. The second level, known as the high-level trigger, consists of a farm of processors running a version of the full event reconstruction software optimized for fast processing, and reduces the event rate to around 1\unit{kHz} before data storage~\cite{Khachatryan:2016bia}.

The silicon tracker used in 2016 measured charged particles within the range $\abs{\eta} < 2.5$. For nonisolated particles of $1 < \pt < 10\GeV$ and $\abs{\eta} < 1.4$, the track resolutions were typically 1.5\% in \pt and 25--90 (45--150)\mum in the transverse (longitudinal) impact parameter~\cite{CMS:2014pgm}. At the start of 2017, a new pixel detector was installed~\cite{Phase1Pixel}; the upgraded tracker measured particles up to $\abs{\eta} = 3.0$ with typical resolutions of 1.5\% in \pt and 20--75\mum in the transverse impact parameter~\cite{DP-2020-049} for nonisolated particles of $1 < \pt < 10\GeV$.

In the barrel section of the ECAL, an energy resolution of about 1\% is achieved for unconverted or late-converting photons in the tens of \GeV energy range. The energy resolution of the remaining barrel photons is about 1.3\% up to $\abs{\eta} = 1$, changing to about 2.5\% at $\abs{\eta} = 1.4$. In the endcaps, the energy resolution is about 2.5\% for unconverted or late-converting photons, and between 3 and 4\% for the other ones~\cite{CMS:2015myp}. Electrons mimicking photons are rejected by requiring the absence of charged particle tracks, with a hit in the innermost layer of the pixel detector not matched to a reconstructed conversion vertex, pointing to the photon cluster in the ECAL~\cite{CMS:2015myp}. The photon energy in simulation is corrected with a multivariate regression to match the energy resolution observed in data~\cite{CMS:2020uim}.

The efficiency to reconstruct and identify muons is greater than 96\%. Matching muons to tracks measured in the silicon tracker results in a relative transverse momentum resolution, for muons with \pt up to 100\GeV, of 1\% in the barrel and 3\% in the endcaps~\cite{CMS:2018rym}.
The simulated muons are calibrated to match the momentum scale and resolution observed in $\PZ\to\PGm\PGm$ data.

The event reconstruction is performed with a particle-flow (PF) algorithm~\cite{Sirunyan:2017ulk}, which uses an optimized combination of information from the various elements of the CMS detector to reconstruct and identify individual particles produced in each collision. The algorithm identifies each reconstructed particle either as an electron, a muon, a photon, or a charged or neutral hadron.
The PF candidates are then clustered into jets using the anti-\kt algorithm~\cite{Cacciari:2008gp} implemented in the \FASTJET package~\cite{Cacciari:2011ma,Cacciari:2008gn} with a distance parameter $R=0.4$.
The charged component of the pileup contribution is removed from jets by applying the charged-hadron subtraction (CHS) mitigation algorithm~\cite{Sirunyan:2020foa}.
Jets that originate from \cPqb quarks are identified with the \textsc{DeepCSV} \cPqb tagging algorithm~\cite{CMS:2017wtu}, which uses the tracks and secondary vertices associated with the jets as inputs to a neural network. The working point considered in this search provides a \cPqb jet tagging efficiency of 70\%, and a misidentification rate in a sample of quark and gluon jets of about 1\%~\cite{CMS:2017wtu}.

\section{Data and simulated samples}\label{sec:mcsimulation}

The data used in this search have been collected with a trigger algorithm that selects events with at least one muon with $\pt > 17\GeV$ and one photon with transverse energy $\et > 30\GeV$. This trigger collected data with an integrated luminosity of 123\fbinv.

The $\PZ/\PH\to\PGyPnS\PGg$ signal Monte Carlo (MC) events are simulated at leading order (LO) accuracy in QCD with the \MGvATNLO~v2.6.5 matrix element generator~\cite{MadGraph}, directly including the spin correlations~\cite{PPHEP} using a model based on Refs.~\cite{Konig:2015qat,Bodwin:2017pzj} instead of reweighting the simulated events as in previous analyses~\cite{CMS:2018gcm,ATLAS:2018xfc,ATLAS:2022rej}.
Whereas the $\PZ$ boson production is simulated through quark-antiquark annihilation ($\qqbar\to\PZ$), the simulation of the \PH boson production is performed separately for each production mode, namely through gluon-fusion ($\Pg\Pg$F), vector-boson fusion (VBF), vector boson associated production ($\PW\PH$ and $\PZ\PH$), and bottom- and top-quark associated production ($\bbbar\PH$ and $\ttbar\PH$). The Higgs boson mass is set to 125\GeV.
For each boson and production mode, two different samples are generated, with the \PZ and \PH bosons decaying to $\PJGy\PGg$ or $\PGyP{2S}\PGg$, where the $\PGyPnS$ decays to a pair of muons.

Cascade decays $\PGyP{2S}\to\PJGy+\PX\to\PGm\PGm+\PX$, even though they provide a nonnegligible number of events compared with the direct decays $\PGyP{2S}\to\PGm\PGm$, are not considered in this analysis because the momentum carried away by the additional particles ($\PX$) makes their invariant mass \mmmg nonresonant and shifts it towards smaller \mmmg values, not contributing to the peak of the expected signal.

The \PZ boson cross section is calculated at next-to-next-to-LO (NNLO) using \FEWZ~v3.1~\cite{Li:2012wna}. The Higgs boson cross section is normalized to a theoretical prediction at up to next-to-NNLO~\cite{LHCHiggsCrossSectionWorkingGroup:2016ypw}. The branching fractions used are listed in Table~\ref{tab:br}. The \PZ boson produced through VBF has a cross section that is several orders of magnitude smaller than the \qqbar production mechanism, and is therefore considered negligible.

The SM electroweak backgrounds include resonant processes that produce a pair of muons and a photon, which consist of \PZ and \PH bosons that decay through quark loops without the intermediate resonant \ccbar state, referred to as ``Dalitz decays'', and \PZ boson decays to a pair of muons with a final-state radiation photon. Both processes are generated at next-to-LO (NLO) in QCD with the \MGvATNLO event generator. The decay of a Higgs boson into two muons with a photon radiated from one of the muons represents another potential source of background, but its contribution is negligible.

In all simulated samples, parton showering and hadronization processes are simulated by interfacing the event generators with~\PYTHIA~8.230 with the CP5 underlying event tune~\cite{CMS:2019csb}.
The NNPDF3.1 NNLO~\cite{2013290,BALL2012153} parton distribution functions (PDFs) are used to model the momentum distribution of the colliding partons inside the protons. The generated events include additional proton-proton interactions within the same or adjacent bunch crossings (pileup) with a distribution that matches that observed in the data.
The response of the CMS detector to the generated particles is modeled using the \GEANTfour toolkit~\cite{Agostinelli:2002hh}. Simulated events are reconstructed with the same algorithms used for data.

\section{Event selection}\label{sec:reconstruction}

Events are required to contain at least two reconstructed muons with $\pt > 18$ and 5\GeV, respectively, fulfilling the \emph{medium} identification requirement~\cite{CMS:2018rym}, and at least one photon with $\pt > 32\GeV$ passing a multivariate identification criteria that is 80\% efficient for signal photons~\cite{CMS:2020uim}. Both the muons and the photon are required to be isolated from other hadronic activity in the event. If more than two muons are present in the same event, the opposite-sign pair with the smallest $\Delta R = \sqrt{\smash[b]{(\Delta\eta)^2+(\Delta\phi)^2}}$ distance is chosen.

The observable used for signal extraction in this search is the invariant mass \mmmg. Events are separated into three regions based on the invariant mass \mmm: (i) events with $3.0 < \mmm < 3.2\GeV$ are considered to be \PJGy candidates and belong to a signal region (SR) called SR1; (ii) events with $3.60 < \mmm < 3.75\GeV$ are assumed to originate from a \PGyP{2S} state and belong to a separate signal region SR2; and (iii) events that do not belong to either SR1 or SR2, but satisfy $2 < \mmm < 8\GeV$, are a control region (CR) that is used to constrain the resonant \mmmg background.

Depending on the signal process, additional categories are defined to exploit distinct physical features. Three categories maximize the sensitivity to the main Higgs boson production modes: (i) the VBF category contains events with at least one pair of jets with $\pt > 30 \, (50)\GeV$ in the central (forward) regions of the detector with an invariant mass larger than 350\GeV; (ii) the heavy-flavor (HF) category contains events with at least one \PQb tagged jet with $\pt > 30\GeV$ and $\abs{\eta}<2.4$ and targets the Higgs boson production in association with bottom and top quarks; (iii) events that do not belong to the previous categories constitute the ggF category. The data set we used does not contain enough events for a dedicated category targeting leptonic decays of the \PW or \PZ boson.

The same categorization is not applicable to the \PZ boson search since the \PZ boson is predominantly produced through \qqbar annihilation. Instead, the \PZ, \PGyPnS, and \PGg spin correlations are used to build a multivariate likelihood discriminator (LD) to separate the $\PZ \to \PJGy\PGg$ signal from the main background.
The input variables of the discriminator are: the cosine of the angle between the momentum directions of the \PJGy boson and of the beam in the rest frame of the \PZ boson ($\cos\theta^*$),
the angle between the momentum directions of the positively charged muon and the \PJGy in the rest frame of the \PJGy ($\cos\theta_1$), which is sensitive to the transverse polarization of the \PJGy along its momentum direction, and the angle between the decay plane of the two muons and the plane of the \PJGy and the photon ($\phi_1$).
Since these variables are uncorrelated with each other and from \mmm and \mmmg, a selection on the discriminator output does not alter the shape of the \mmmg distribution. The LD is used to define two categories, denoted as high-purity (Z-HP) and low-purity (Z-LP) categories, which contain events with an LD score larger or smaller than 0.5, respectively. The discriminator is applied to only the $\PZ\to\PJGy\PGg$ category to separate the signal and the background.
In the $\PH\to\PJGy\PGg$ ggF region, most of the discriminating power comes from the $\abs{\cos\theta^*}$ variable.
Thus, this variable is used in place of the LD to define high-purity (ggF-HP) and low-purity (ggF-LP) categories, containing events with a $\abs{\cos\theta^*}$ value larger or smaller than 0.5, respectively.
Because of the limited number of events in the other signal regions, the LD categorization or the angular selections are not applied.

The selections used to divide events into the eight signal regions and the control region are summarized in Table~\ref{tab:selections}.

\begin{table*}[!htp]\centering
    \topcaption{Summary of the selections of the seven signal regions (SRs) and the control region (CR). The SR1 category requires an \mmm value compatible with the \PJGy mass hypothesis, and is further divided depending on the value of the LD variable (Z-HP, Z-LP), the Higgs boson production mode (ggF, VBF, HF), and the value of the $\cos\theta^*$ variable (ggF-HP, ggF-LP). The SR2 category requires a \mmm value compatible with the \PGyP{2S} mass hypothesis and is not further categorized. The CR category is made orthogonal to the SRs by vetoing \mmm values compatible with the \PGyPnS mass hypotheses.}
    \label{tab:selections}
    \cmsTableMaybe{
        \begin{tabular}{lcccccccc}

                                     & \multicolumn{6}{c}{SR1}                                & SR2           & CR            \\
            \cline{2-7}
                                     & Z-HP   & Z-LP    & ggF-HP  & ggF-LP  & VBF & HF        &               &               \\
            \hline
            $\pt(\PGm_1)$            & \multicolumn{8}{c}{$>$18\GeV}                                                          \\
            $\pt(\PGm_2)$            & \multicolumn{8}{c}{$>$5\GeV}                                                           \\
            $\pt(\PGg)$              & \multicolumn{8}{c}{$>$32\GeV}                                                          \\
            $\abs{\eta(\PGm_{1,2})}$ & \multicolumn{8}{c}{$<$2.4}                                                             \\
            $\abs{\eta^{\PGg}}$      & \multicolumn{8}{c}{$<$2.5, excluding ECAL transition region}                           \\
            [\cmsTabSkip]
                                     &        &         &         &         &     &           &               & [2.0, 3.0],   \\
            \mmm (\GeVns)            & \multicolumn{6}{c}{[3.0, 3.2]}                         & [3.6, 3.75]   & [3.2, 3.6],   \\
                                     &        &         &         &         &     &           &               & [3.75, 8.0]   \\
            [\cmsTabSkip]
            $\abs{\cos\theta^{*}}$   & \NA    & \NA     & $>$0.5  & $<$0.5  & \NA & \NA       & \NA           & \NA           \\
            LD                       & $>$0.5 & $<$0.5  & \NA     & \NA     & \NA & \NA       & \NA           & \NA           \\
            \PQb jets                & \NA    & \NA     & 0       & 0       & 0   & ${\geq}1$ & \NA           & \NA           \\
            VBF candidates           & \multirow{2}{*}{\NA} & \multirow{2}{*}{\NA}  & \multirow{2}{*}{0}  & \multirow{2}{*}{0}  & \multirow{2}{*}{${\geq}1$} & \multirow{2}{*}{0} & \multirow{2}{*}{\NA} & \multirow{2}{*}{\NA}  \\
            ($m_\text{jj}>350\GeV$)  & \\
        \end{tabular}
    }
\end{table*}

\section{Signal and background modeling}\label{sec:bkg}

Backgrounds are separated into two classes. The electroweak-induced backgrounds are composed of Dalitz decays of the Higgs boson and $\PZ\to\PGm\PGm$ events with final-state radiation, and are modeled with MC simulation. The QCD multijet production, which does not exhibit a resonant shape in \mmmg, is estimated from data with parametric functions.

The resonant backgrounds and the signals appear as resonances in the \mmmg distribution and are modeled with double-sided Crystal Ball (DSCB) functions~\cite{DSCBRef}, which consist of a Gaussian core and two power-law functions that describe the lower and upper tails, below and above a certain threshold.
An alternative Voigt distribution, which consists of a convolution of a Gaussian with a Cauchy-Lorentzian, is also evaluated.
Both functions are able to describe the resonant shape, but the DSCB is chosen because of its better modeling of the lower tails.
The Gaussian core of the DSCB fitted to the \PH boson signal is dominated by the measurement resolution and has a relative width of 1\%.
The \PZ boson signal width, instead, is dominated by the natural width of the \PZ resonance.

The nonresonant background is modeled separately for each category directly from data using parametric functions to describe the \mmmg distribution. Several choices of function families are considered in the statistical analysis and, for each family, models with different numbers of parameters are included. These functions are: power-law and exponential with a number of parameters between 1 and 4; Bernstein polynomials with a number of parameters between 3 and 5.
For each function family, a procedure based on the Fisher F-test~\cite{Ftest} starts from the simplest functional form and checks iteratively, using a confidence level of 95\%, if additional parameters are needed to model the background distributions.

\begin{figure*}[!htb]\centering
  \includegraphics[width=0.45\textwidth]{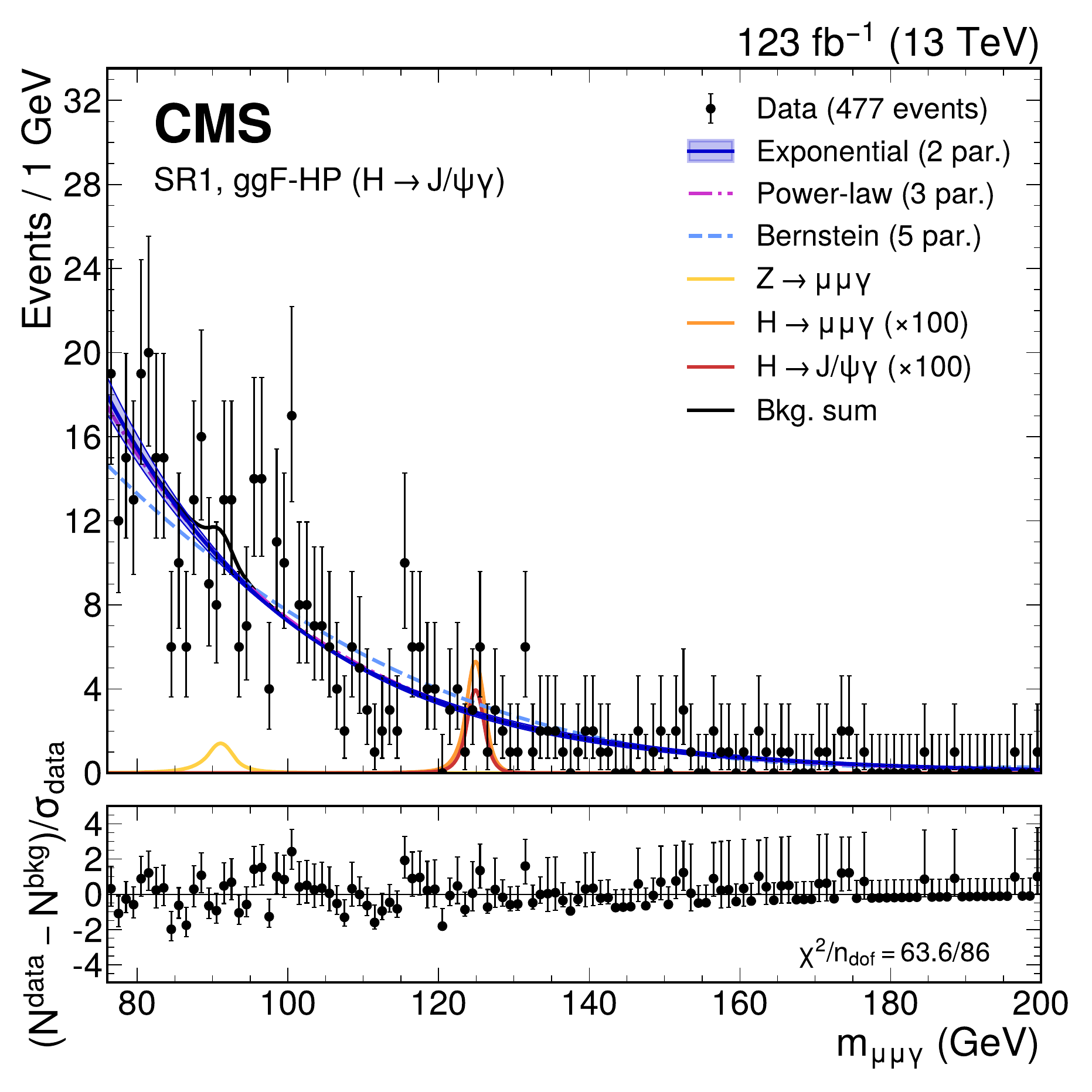}
  \includegraphics[width=0.45\textwidth]{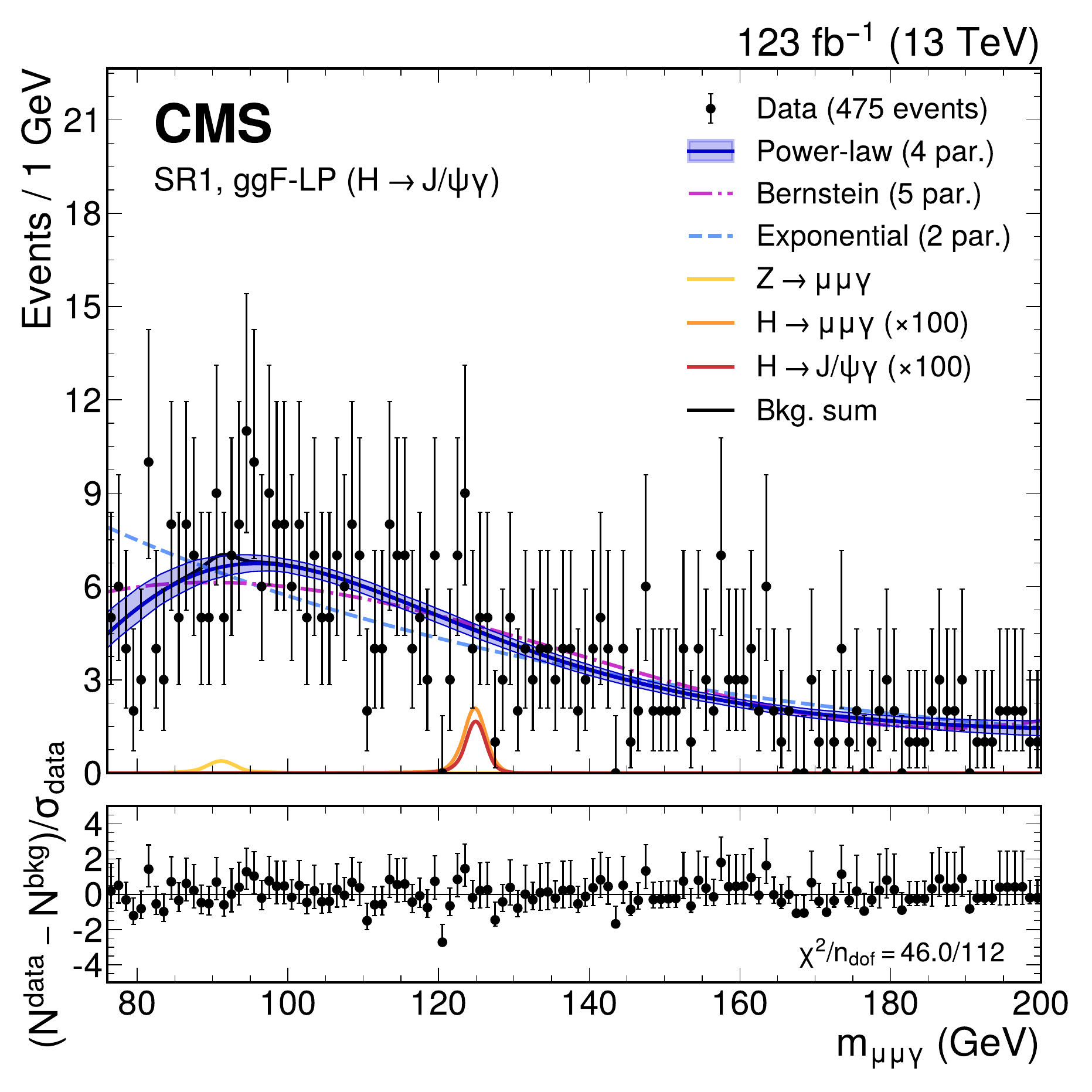} \\
  \includegraphics[width=0.45\textwidth]{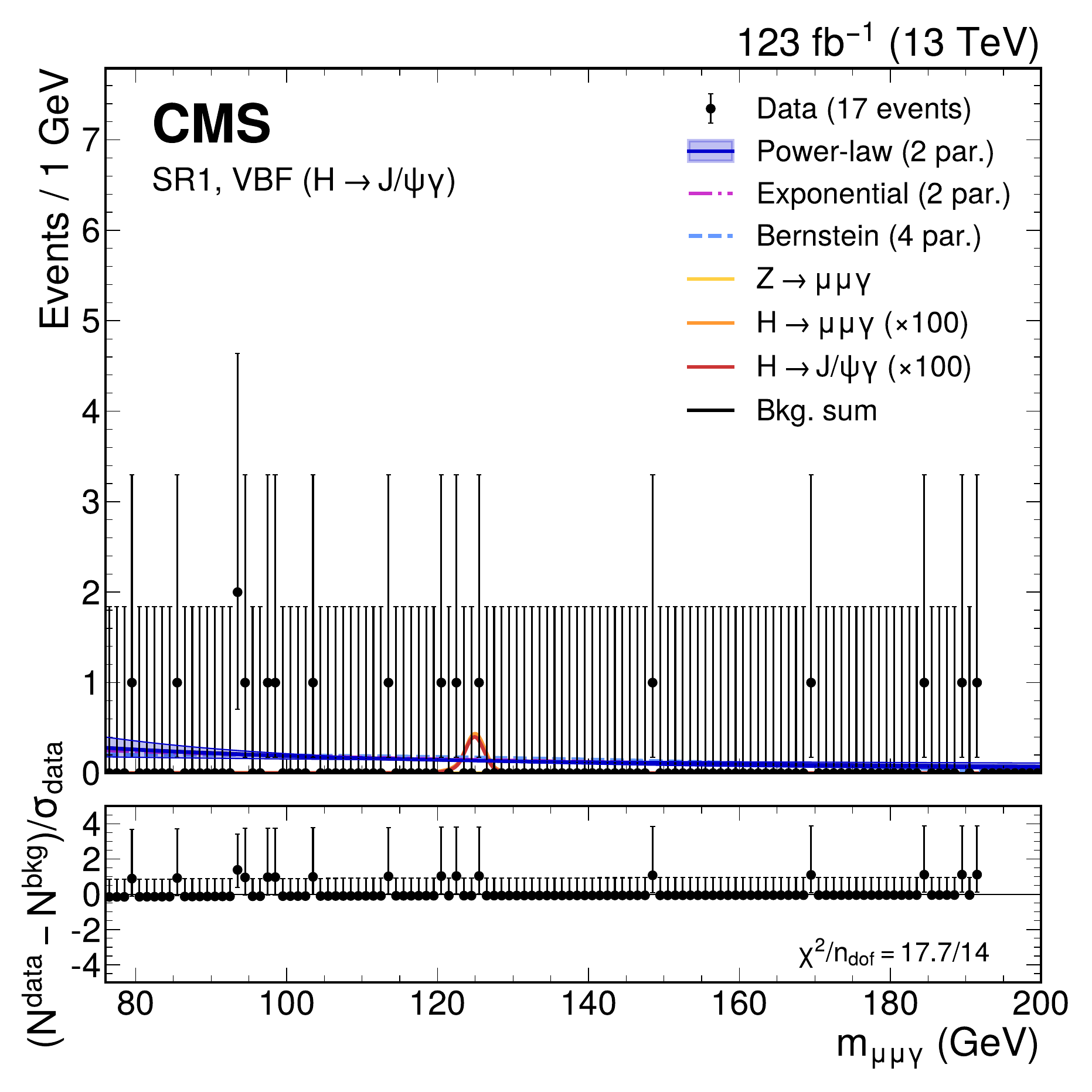}
  \includegraphics[width=0.45\textwidth]{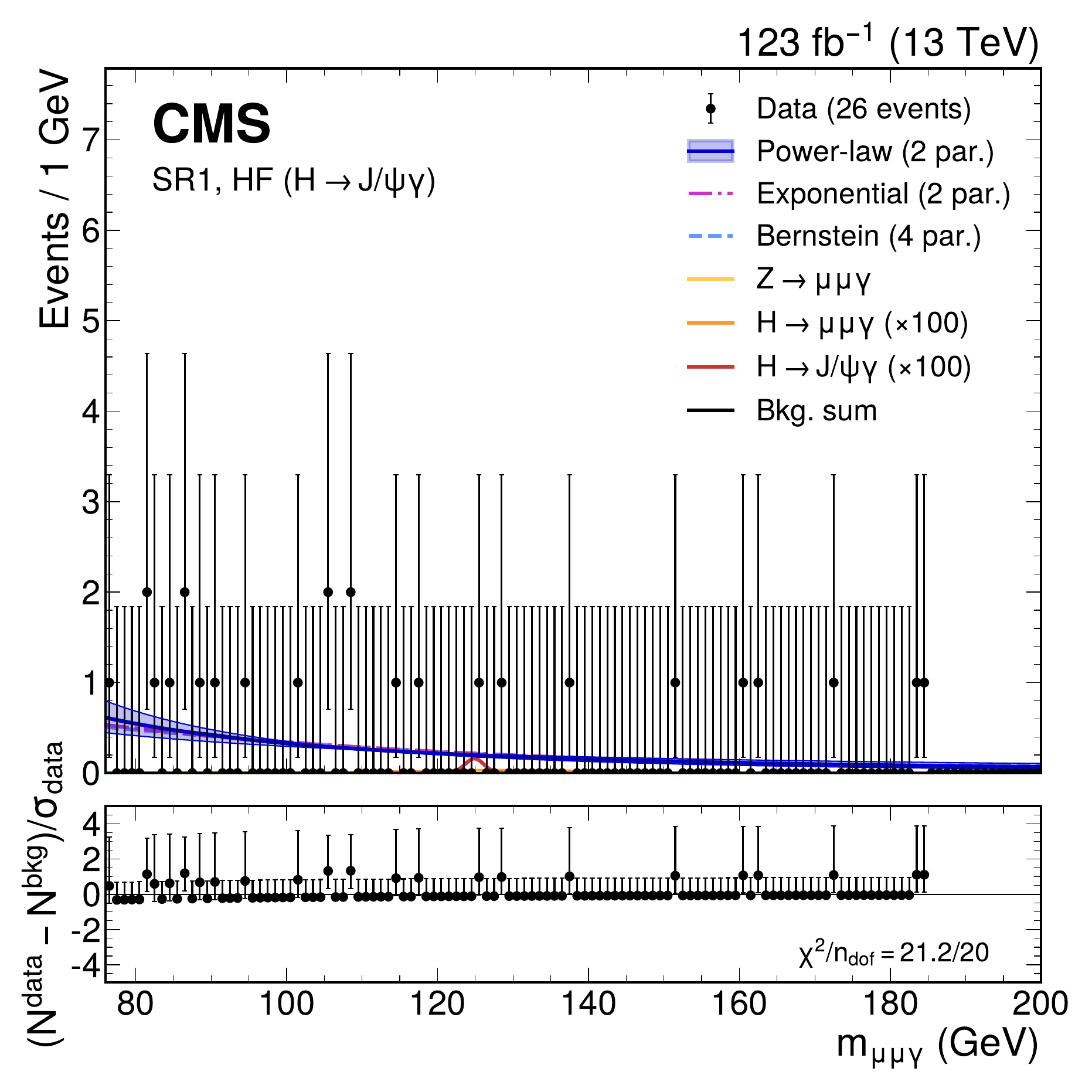}
  \caption{The upper panels show the background-only fit using the power-law, exponential, and Bernstein polynomial functions with the optimal number of parameters determined with the F-test, and ranked according to the fit $\chi^2$. The resonant background contributions are added with the normalization fixed to the SM expectation. The vertical error bars on the data points show the statistical uncertainty. The bottom panels show the pulls for the fit result for the model with lowest $\chi^2$. Upper left: $\PH\to\PJGy\PGg$ ggF-HP category. Upper right: $\PH\to\PJGy\PGg$ ggF-LP category. Lower left: $\PH\to\PJGy\PGg$ VBF category. Lower right: $\PH\to\PJGy\PGg$ HF category.}\label{fig:bkg_fits_H}
\end{figure*}

\begin{figure*}[!htb]\centering
  \includegraphics[width=0.45\textwidth]{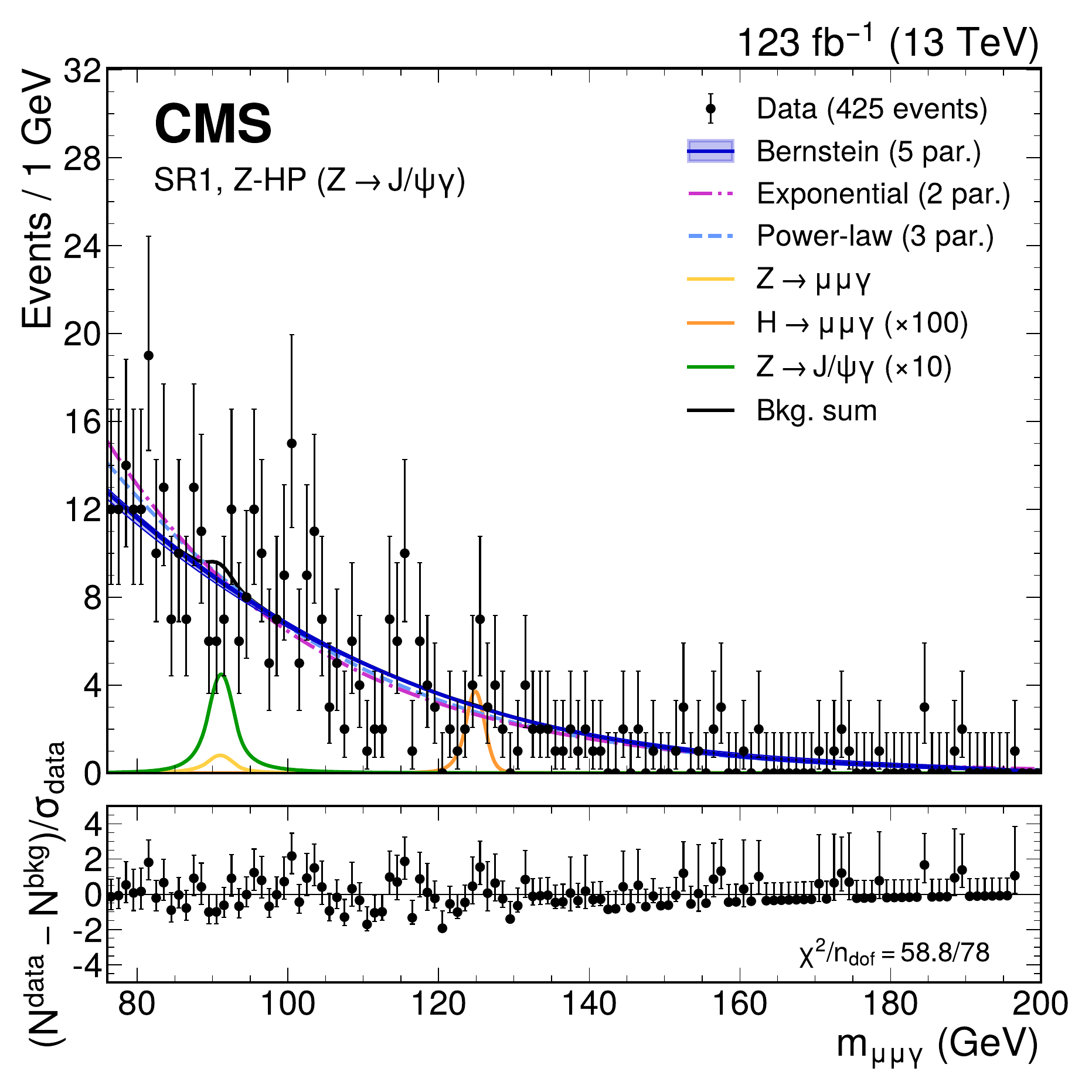}
  \includegraphics[width=0.45\textwidth]{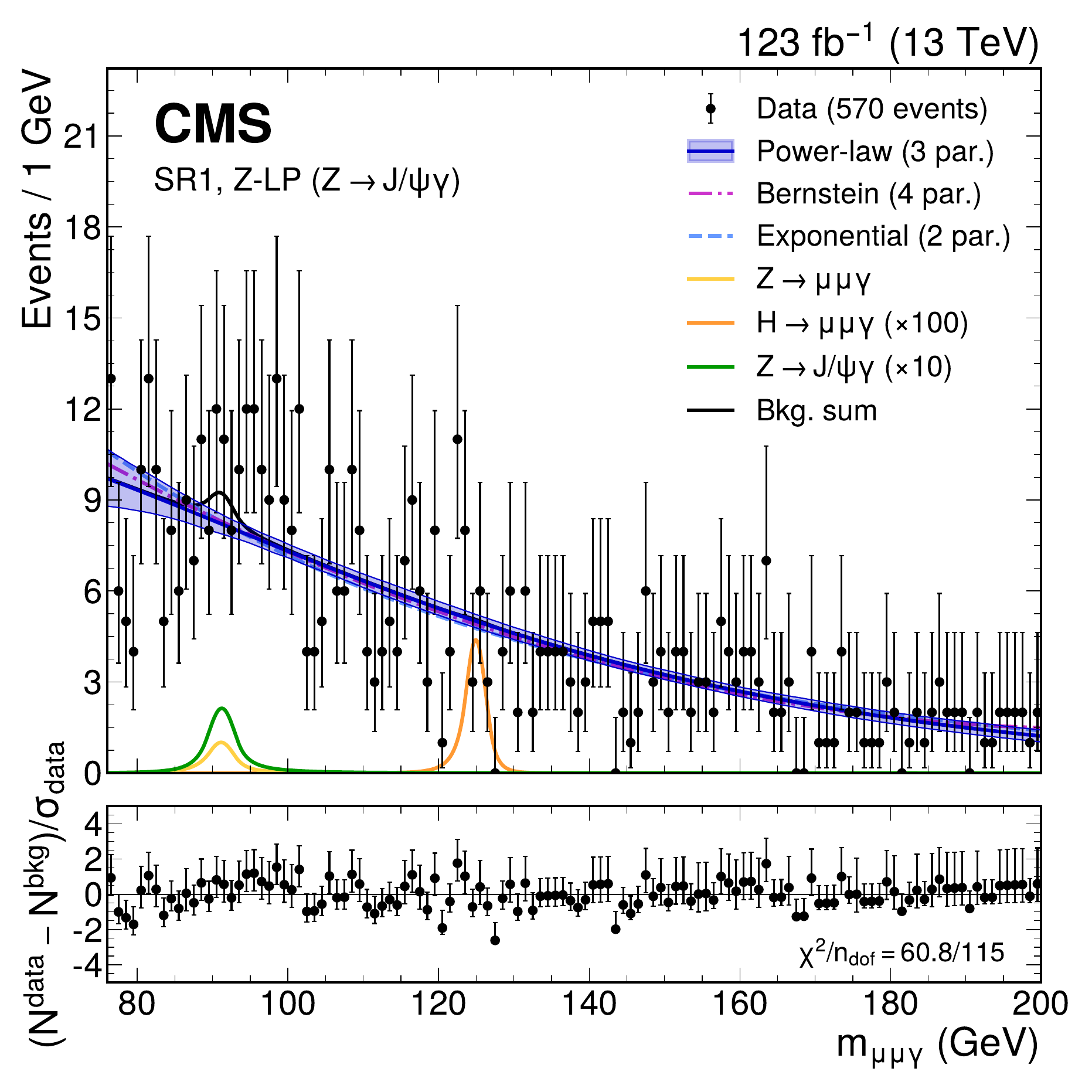} \\
  \includegraphics[width=0.45\textwidth]{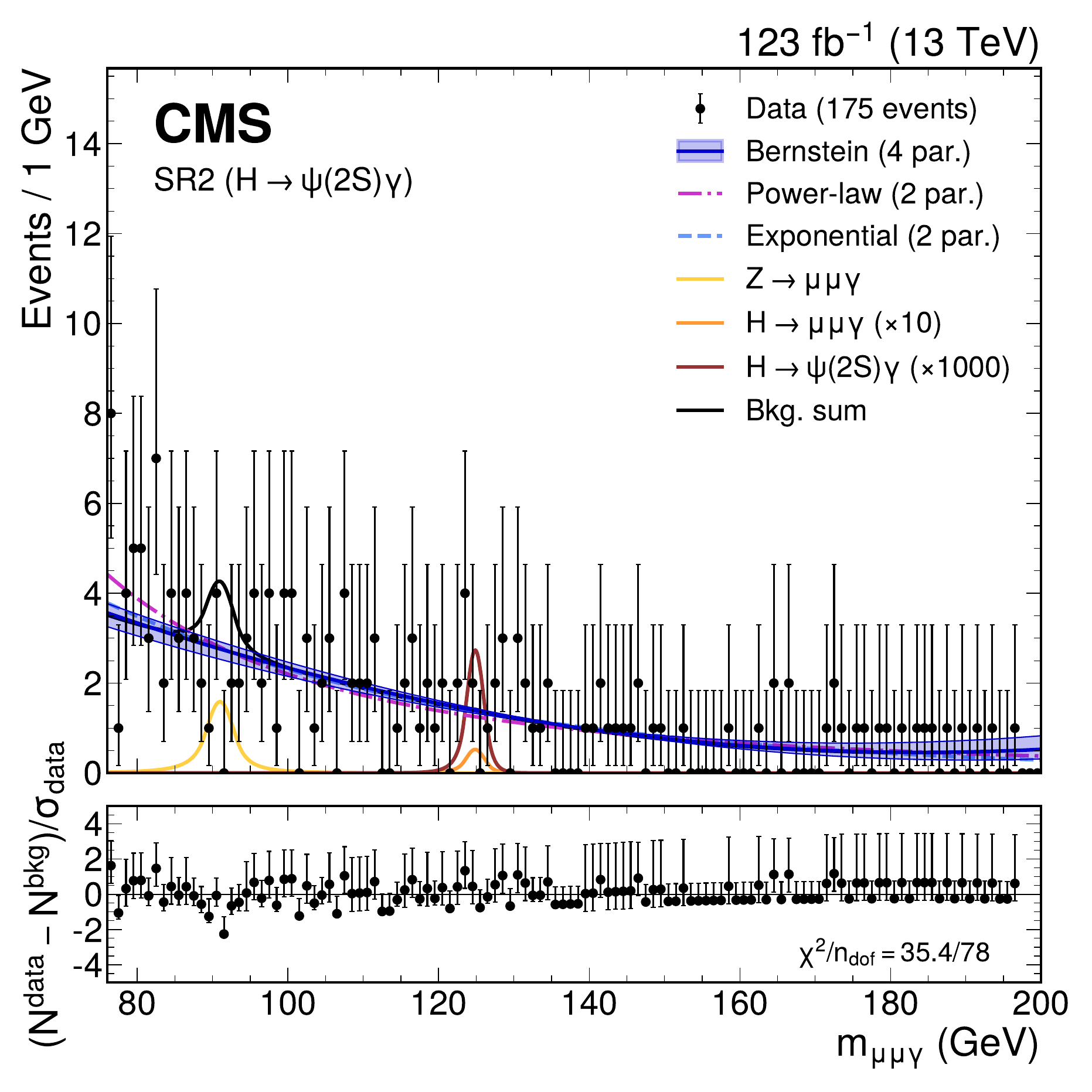}
  \includegraphics[width=0.45\textwidth]{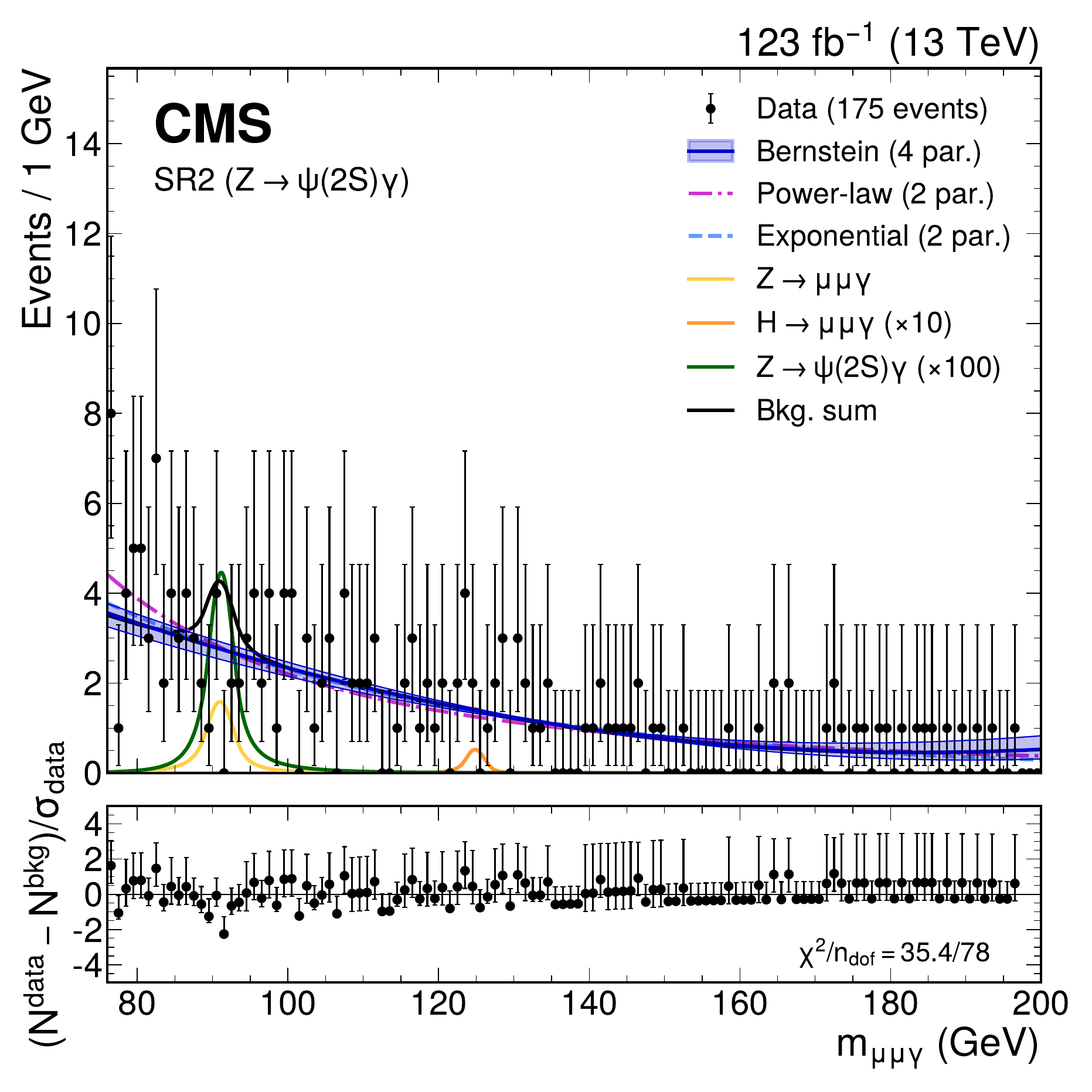}
  \caption{The upper panels show the background-only fit using the power-law, exponential, and Bernstein polynomial functions with the optimal number of parameters determined with the F-test, and ranked according to the fit $\chi^2$. The resonant background contributions are added with the normalization fixed to the SM expectation. The vertical error bars on the data points show the statistical uncertainty. The bottom panels show the pulls for the fit result for the model with lowest $\chi^2$. Upper left: $\PZ\to\PJGy\PGg$ HP category. Upper right: $\PZ\to\PJGy\PGg$ LP category. Lower left: $\PH\to\PGyP{2S}\PGg$ category. Lower right: $\PZ\to\PGyP{2S}\PGg$ category.}\label{fig:bkg_fits_Z}
\end{figure*}

\begin{figure}[!htb]\centering
  \includegraphics[width=\cmsFigWidth]{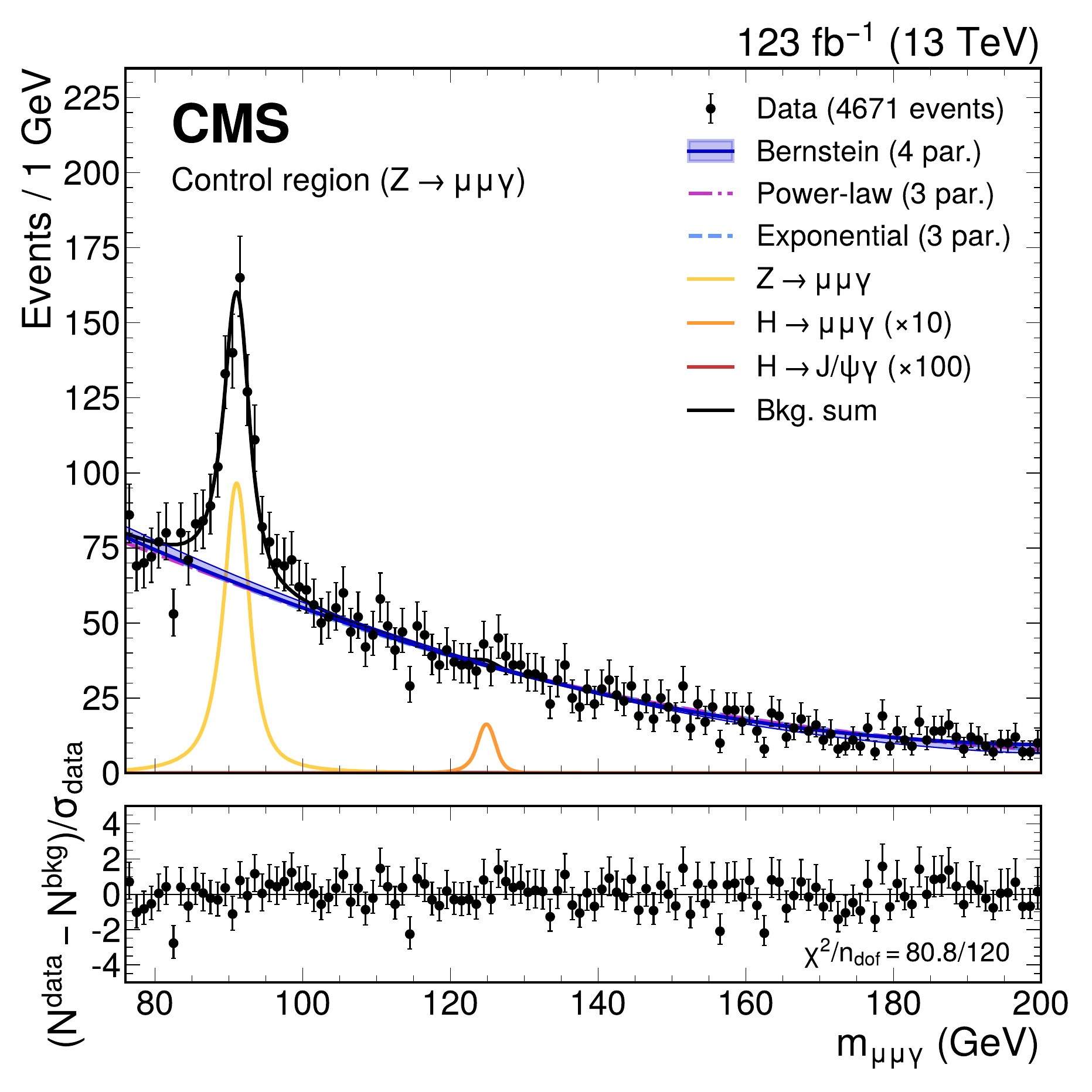}
  \caption{The upper panel shows the results of the main background fit in the CR, using the power-law, exponential, and Bernstein polynomial functions with the optimal number of parameters determined with the F-test, and ranked according to the fit $\chi^2$. The vertical error bars on the data points show the statistical uncertainty. The bottom panel shows the pulls for the fit result for the model with lowest $\chi^2$.}\label{fig:bkg_fits_CR}
\end{figure}

The measured \mmmg distributions are well modeled by the background parametric functions, as shown in Figs.~\ref{fig:bkg_fits_H}--\ref{fig:bkg_fits_CR}. The width of the mass bins is chosen according to the experimental mass resolution.

\section{Systematic uncertainties}\label{sec:sys}

Systematic uncertainties include the modeling of the backgrounds, the detector simulation, the integrated luminosity, and the theoretical prediction of the signal production mechanism.

The uncertainty related to the choice of function used to model the QCD multijet background is addressed by introducing the \textit{discrete profiling} method~\cite{Dauncey:2014xga}.
This method considers all function families listed in Section~\ref{sec:bkg}, with the number of free parameters determined with an F-test, and fitted to the appropriate \mmmg distribution in each category by minimizing the negative logarithm of the likelihood multiplied by two (2NLL).
The functions with resulting fit \textit{p}-values smaller than 0.05 (and that, hence, do not fit the data sufficiently well) are discarded.
A unitary penalty term is added to the 2NLL to account for the number of floating parameters in each candidate function.
The envelope of the 2NLL profiles for each function considered, as a function of the signal strength, is then used to determine the best-fit value and the corresponding uncertainty in the parameter of interest.
The outcome of this procedure is that the selection of the background function is treated as a discrete nuisance parameter in the fit to account for the uncertainty associated with the arbitrariness of this choice.

An additional uncertainty is introduced to cover potential biases in the choice of the function used to model the signal and the resonant backgrounds. An alternate Voigtian function is used and the difference in the fitted signal strength, which is always smaller than 1\%, is considered as an uncertainty.

The experimental systematic uncertainties affecting the signal normalization include the reconstruction and selection efficiency of the muons and the photon, amounting to 2.0 and 1.5\%, respectively. The muon momentum scale and resolution calibration uncertainties are propagated to the \mmmg distribution, and have an effect of 0.05 and 0.80\% on the Gaussian mean and width used to model the simulated samples, respectively. Similarly, the photon energy scale and photon energy resolution result in a 3.4 and 1.8\% uncertainty in the Gaussian width, respectively.
The jet energy scale and resolution uncertainties affect the Higgs boson normalization in the VBF category by 2.9 and 1.0\%, respectively, while the jet \PQb tagging efficiency uncertainty is 5.4\% in the HF category. The effect of these uncertainties is negligible elsewhere.
The uncertainty in the efficiency of the trigger is less than 3.2\%.
During the 2016 and 2017 data-taking periods, a gradual shift in the timing of the inputs of the ECAL L1 trigger in the $\abs{\eta} > 2.0$ region caused a specific trigger inefficiency. For events containing a photon (a jet) with \pt larger than $\approx$50\GeV ($\approx$100\GeV) in the $2.5 < \abs{\eta} < 3.0$ region, the efficiency loss is $\approx$10--20\%, depending on \pt, $\eta$, and time. Correction factors were computed from data and applied to the acceptance evaluated with simulation.
Other uncertainties, such as the pileup contributions (0.1\%), and the measurement of the integrated luminosity (1.2\% in 2016~\cite{CMS:2021xjt}, 2.3\% in 2017~\cite{CMS:2018elu}, and 2.5\% in 2018~\cite{CMS:2019jhq}) affect only the normalization of the simulated samples and have negligible effects on their \mmmg distributions.

Systematic uncertainties in the theoretical prediction arise from the choice of PDFs~\cite{Ball:2014uwa,Butterworth:2015oua}, leading to a 2--3\% effect on the signal cross section for the \PZ and \PH boson signals. The factorization and renormalization scale uncertainties are estimated by varying the scales up and down by a factor of 2, both simultaneously and independently, using the maximum obtained value. The resulting effect is a variation of 3--6\% of the signal cross section.

A summary of the systematic uncertainties is provided in Table~\ref{tab:systematics}.

\begin{table*}[htp]\centering
    \topcaption{Sources and types of systematic uncertainties described in Section~\ref{sec:sys}. Normalization (norm.) uncertainties yield a variation in the number of events equal to the reported value. Other uncertainties also vary the shape of the mean and the width of the resonant background and signal \mmmg distributions by the indicated amount. Uncertainties in the same line are treated as correlated.}\label{tab:systematics}
    \cmsTable{
        \begin{tabular}{lcccccc}
            Source                                                      &
            Type                                                        &
            \multicolumn{3}{c}{Backgrounds}                             &
            \multicolumn{2}{c}{Signals}                                                                                                     \\
            \hline
            &                                                           &
            QCD                                                         &
            $\PZ\to\PGm\PGm\PGg$                                        &
            $\PH\to\PGm\PGm\PGg$                                        &
            $\PZ\to\PGyPnS\PGg$                                         &
            $\PH\to\PGyPnS\PGg$                                                                                                           \\
            \hline
            Integrated luminosity                                       & norm.       &   \NA         &   \NA             &   1.6\%     &   1.6\%     &   1.6\%     \\
            Fact.\ and renorm.\ scales                                  & norm.       &   \NA         &   \NA             &   3.6\%     &   3.5\%     &   5.6\%     \\
            PDF                                                         & norm.       &   \NA         &   \NA             &   1.8\%     &   1.7\%     &   3.2\%     \\
            $\PH\to\PGm\PGm\PGg$ branching ratio                        & norm.       &   \NA         &   \NA             &   5.3\%     &   \NA       &   \NA       \\
            Pileup modeling                                             & norm.       &   \NA         &   \NA             &   0.7\%     &   0.9\%     &   0.8\%     \\
            Trigger efficiency                                          & norm.       &   \NA         &   \NA             &   3.5\%     &   3.5\%     &   3.5\%     \\
            Trigger timing shift                                        & norm.       &   \NA         &   \NA             &   0.0\%     &   0.6\%     &   2.7\%     \\
            Muon identification/isolation                               & norm.       &   \NA         &   \NA             &   1.6\%     &   2.0\%     &   1.4\%     \\
            Photon identification                                       & norm.       &   \NA         &   \NA             &   1.6\%     &   1.5\%     &   1.5\%     \\
            Electron veto                                               & norm.       &   \NA         &   \NA             &   1.0\%     &   1.0\%     &   1.0\%     \\
            Jet energy scale (VBF category)                             & norm.       &   \NA         &   \NA             &   2.9\%     &   \NA       &   2.9\%     \\
            Jet energy resolution (VBF category)                        & norm.       &   \NA         &   \NA             &   1.0\%     &   \NA       &   1.0\%     \\
            Jet \PQb tagging (HF category)                              & norm.       &   \NA         &   \NA             &   5.4\%     &   \NA       &   5.4\%     \\
            QCD background model                                        & shape       &   discrete    &   \NA             &   \NA       &   \NA       &   \NA       \\
            Resonant bkg. model                                         & norm.       &   \NA         &   1\%             &   \NA       &   \NA       &   \NA       \\
            Signal model                                                & norm.       &   \NA         &   \NA             &   1\%       &   1\%       &   1\%       \\
            Size of MC samples                                          & \mmmg mean  &   \NA         &   0.06\%          &   0.17\%    &   0.06\%    &   0.06\%    \\
            Size of MC samples                                          & \mmmg width &   \NA         &   0.1\%           &   1.3\%     &   0.1\%     &   0.1\%     \\
            \Pgm calibration                                            & \mmmg mean  &   \NA         &   0.05\%          &   0.05\%    &   0.05\%    &   0.05\%    \\
            \Pgm calibration                                            & \mmmg width &   \NA         &   0.8\%           &   0.8\%     &   0.8\%     &   0.8\%     \\
            \PGg energy                                                 & \mmmg mean  &   \NA         &   0.04\%          &   0.04\%    &   0.04\%    &   0.04\%    \\
            \PGg energy                                                 & \mmmg width &   \NA         &   3.4\%           &   3.4\%     &   3.4\%     &   3.4\%     \\
            \PGg resolution                                             & \mmmg width &   \NA         &   1.8\%           &   1.8\%     &   1.8\%     &   1.8\%     \\
        \end{tabular}
    }
\end{table*}

\section{Results and interpretation}\label{sec:res}

In the absence of a significant observed signal, exclusion limits are obtained by performing a background-only fit and a combined signal-plus-background fit to the \mmmg mass distributions, independently for the \PZ and Higgs boson signals, and the \PJGy and \PGyP{2S} hypotheses.

The Higgs and \PZ boson signals are fitted to the \PH and \PZ SRs, respectively. Each fit also includes the CR, which constrains the normalization of the \PZ boson Dalitz backgrounds. The \PH boson Dalitz backgrounds are fixed to the SM background expectation, because the number of selected events is not sufficient to constrain this process.

In the fit, based on a profile likelihood, the parameters and the normalization of the background in each category are left free to float.
The systematic uncertainties in the signal are treated as nuisance parameters profiled in the statistical interpretation~\cite{CMS-NOTE-2011-005}, using log-normal constraints for the normalization uncertainties and Gaussian constraints for the muon and photon energy scale and resolution.
The \PZ boson Dalitz background signal strength, $\mu_{\PZ\to\PGm\PGm\PGg}$, is extracted from a standalone fit to the \mmmg distribution measured for the CR events (shown in Fig.~\ref{fig:bkg_fits_CR}). The result is statistically compatible with the SM expectation: $\mu_{\PZ\to\PGm\PGm\PGg} = 1.18 \pm 0.12$.

Upper limits at 95\% confidence level (\CL) are set using the \CLs criterion~\cite{CLS1,CLS2} in the asymptotic approximation~\cite{Asymptotic} and cross-checked with those obtained using the generation of pseudo-data~\cite{CMS-NOTE-2011-005}. The limits derived with the two methods are in agreement within 5\%.

\begin{figure}[!htb]\centering
  \includegraphics[width=\cmsFigWidth]{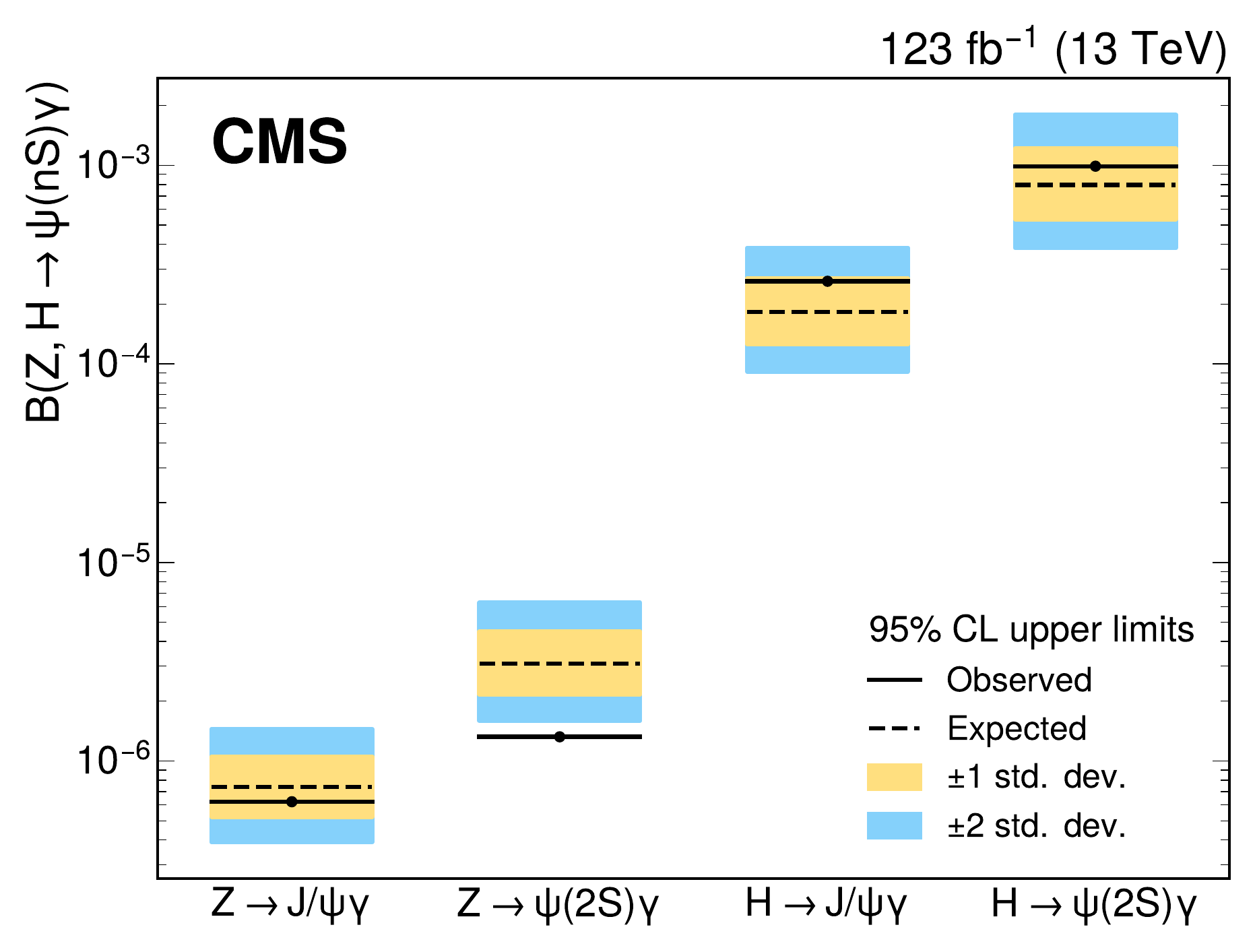}
  \caption{Observed and expected (with ${\pm}1$, ${\pm}2$ standard deviation bands) exclusion limits on the branching fraction of the $(\PH,\PZ) \to \PGyPnS\PGg$ decays.}\label{fig:results}
\end{figure}

The results are summarized in Fig.~\ref{fig:results} and Table~\ref{tab:results}, where the branching fraction and signal strength values for each channel are reported. 
The uncertainties that affect the signal normalization (factorization and renormalization scales, and PDF) are treated differently depending on how the exclusion is presented. When deriving upper limits on $\mathcal{B}$ or $\sigma\,\mathcal{B}$ these uncertainties are not varied in the fit. When placing limits on the signal strength $\mu$, these nuisance parameters are fixed at their best-fit values, in the same manner as the other systematic uncertainties.
No statistically significant excess is observed. A deficit in the data consistent with a downward fluctuation with a significance of 2.2 standard deviations is seen in the $\PZ\to\PGyP{2S}\PGg$ search channel.
The previous results from the CMS and ATLAS Collaborations are also included for comparison in Table~\ref{tab:results}~\cite{CMS:2018gcm,ATLAS:2022rej}.

\begin{table*}[!htp]\centering
    \topcaption{Observed (expected) upper limits at 95\% \CL on the normalized values with respect to the SM expectation, denoted as the signal strength parameter $\mu$, of the product of the cross section $\sigma$ and the branching fraction $\mathcal{B}$ of the $(\PH,\PZ) \to \PGyPnS\PGg$ decays, and the branching fraction, assuming a SM \PZ and \PH boson cross section. The results are compared with previous ones~\cite{CMS:2018gcm,ATLAS:2022rej}.}
    \label{tab:results}
    \renewcommand{\arraystretch}{1.4}
    \cmsTable{
        \begin{tabular}{cccccc}
            \multirow{2}{*}{Process} & \multicolumn{3}{c}{This work (123\fbinv)} & CMS (36\fbinv)~\cite{CMS:2018gcm} & ATLAS (139\fbinv)~\cite{ATLAS:2022rej} \\ \cline{2-4}
                                     & $\mu_\text{obs} (\mu_\text{exp})$         & $\sigma\,\mathcal{B}_\text{obs} (\sigma\,\mathcal{B}_\text{exp}) [\unit{pb}]$                 & $\mathcal{B}_\text{obs} (\mathcal{B}_\text{exp})$                & $\mathcal{B}_\text{obs} (\mathcal{B}_\text{exp})$                & $\mathcal{B}_\text{obs} (\mathcal{B}_\text{exp})$                \\
            \hline
            $\PZ \to \PJGy\PGg$      & $7.2 \, (8.6^{+4.1}_{-2.7})$    & $3.8 \, (4.4^{+1.9}_{-1.3}) \times 10^{-2}$   & $0.6 \, (0.7^{+0.3}_{-0.2}) \times 10^{-6}$ & $1.5 \, (1.7^{+0.7}_{-0.5}) \times 10^{-6}$ & $1.2 \, (0.7^{+0.3}_{-0.2}) \times 10^{-6}$ \\
            $\PZ \to \PGyP{2S}\PGg$  & $29 \, (68^{+36}_{-22})$        & $8 \, (19^{+8}_{-6}) \times 10^{-2}$          & $1.3 \, (3.1^{+1.4}_{-0.9}) \times 10^{-6}$ & \NA & $2.4 \, (3.0^{+1.3}_{-0.8}) \times 10^{-6}$ \\
            $\PH \to \PJGy\PGg$      & $88 \, (62^{+30}_{-19})$        & $1.4 \, (1.0^{+0.5}_{-0.3}) \times 10^{-2}$   & $2.6 \, (1.8^{+0.9}_{-0.6}) \times 10^{-4}$ & $7.6 \, (5.2^{+2.4}_{-1.6}) \times 10^{-4}$ & $2.0 \, (1.8^{+0.8}_{-0.5}) \times 10^{-4}$ \\
            $\PH \to \PGyP{2S}\PGg$  & $970 \, (780^{+420}_{-260})$    & $5.5 \, (4.4^{+2.3}_{-1.5}) \times 10^{-2}$   & $9.9 \, (8.0^{+4.2}_{-2.6}) \times 10^{-4}$ & \NA & $10.5 \, (8.1^{+3.6}_{-2.3}) \times 10^{-4}$ \\
        \end{tabular}
    }
\end{table*}

The upper limits on the $\PH\to\PJGy\PGg$ signal strength are interpreted within the $\kappa$-framework~\cite{Heinemeyer:2013tqa} to constrain the \cPqc quark Yukawa coupling.
The ratios of the signal strengths for the $\PH\to\PJGy\PGg$ and $\PH\to\PGg\PGg$ decays and their SM expectations are approximated as the ratio of their respective partial decay widths, normalized to their SM expectations $\Gamma^{\text{SM}}$.
Writing the partial $\PH\to\PJGy\PGg$ decay width in terms of both direct and indirect amplitudes $\mathcal{A}_{\text{dir}}$ and $\mathcal{A}_{\text{ind}}$ from the diagrams reported in Fig.~\ref{fig:signal}, the ratio $\kc/\kg$ between the coupling modifiers for the \cPqc quark Yukawa coupling and the effective coupling of the Higgs boson to photons can be expressed as
\ifthenelse{\boolean{cms@external}}
{
\begin{multline*}
\begin{aligned}
  \frac{\mu(\PH\to\PJGy\PGg)}{\mu(\PH\to\PGg\PGg)} \approx & \frac{\Gamma(\PH\to\PJGy\PGg)/\Gamma^{\text{SM}}(\PH\to\PJGy\PGg)}{\Gamma(\PH\to\PGg\PGg)/\Gamma^{\text{SM}}(\PH\to\PGg\PGg)} = \\
  = & \frac{|\mathcal{A}_{\text{ind}}\kg + \mathcal{A}_{\text{dir}}\kc|^2}{\kg^{2}} \frac{1}{\Gamma^{\text{SM}}(\PH\to\PJGy\PGg)} = \\
  = & \frac{|\mathcal{A}_{\text{ind}} + \mathcal{A}_{\text{dir}}\kc/\kg|^2}{\Gamma^{\text{SM}}(\PH\to\PJGy\PGg)}.
\end{aligned}
\end{multline*}
}
{
\begin{equation*}
\begin{aligned}
  \frac{\mu(\PH\to\PJGy\PGg)}{\mu(\PH\to\PGg\PGg)}  & \approx \frac{\Gamma(\PH\to\PJGy\PGg)/\Gamma^{\text{SM}}(\PH\to\PJGy\PGg)}{\Gamma(\PH\to\PGg\PGg)/\Gamma^{\text{SM}}(\PH\to\PGg\PGg)} = \\
  & = \frac{|\mathcal{A}_{\text{ind}}\kg + \mathcal{A}_{\text{dir}}\kc|^2}{\kg^{2}} \frac{1}{\Gamma^{\text{SM}}(\PH\to\PJGy\PGg)} = \frac{|\mathcal{A}_{\text{ind}} + \mathcal{A}_{\text{dir}}\kc/\kg|^2}{\Gamma^{\text{SM}}(\PH\to\PJGy\PGg)}.
\end{aligned}
\end{equation*}
}
In this interpretation, the signal strength for $\PH\to\PGg\PGg$ comes from Ref.~\cite{Sirunyan2021:diphoton}, whereas the values for the direct and indirect amplitudes for $\PH\to\PJGy\PGg$ are extracted from Ref.~\cite{Brambilla:2019fmu}.
This interpretation follows the methodology used by the ATLAS Collaboration~\cite{ATLAS:2022rej}, which is based on Refs.~\cite{Brambilla:2019fmu,PhysRevD.92.033016}.
The observed constraint on the $\kc/\kg$ ratio at 95\% \CL corresponds to the interval $(-157,\, +199)$, whereas the expected interval from pseudo-experiments is $(-121,\, +161)$. These results are similar to those reported by ATLAS~\cite{ATLAS:2022rej}.

An alternative interpretation is considered. Under the pure SM assumption, $\kg = 1$ and consequently $\mu(\PH\to\PGg\PGg) = 1$. Thus, the previous equation simplifies to
\begin{equation*}
  \mu(\PH\to\PJGy\PGg) \approx \frac{|\mathcal{A}_{\text{ind}} + \mathcal{A}_{\text{dir}}\kc|^{2}}{\Gamma^{\text{SM}}(\PH\to\PJGy\PGg)}.
\end{equation*}
In this interpretation, the constraint is applied to \kc directly. The observed and expected intervals at 95\% \CL are $(-166,\, +208)$ and $(-128,\, +169)$, respectively.

With respect to the previous CMS search~\cite{CMS:2018gcm}, the sensitivity of this analysis shows an improvement that exceeds the expectations from a simple increase in the integrated luminosity because of the optimization of the event selection, including the angular variables of the decays, the categorization of the Higgs boson depending on the production mode, and the adoption of a more advanced statistical treatment represented by the discrete profiling method, described in Section~\ref{sec:sys}.
When compared with the 2016 CMS search updated with the larger data set, the selections optimized to leverage the unique features of each signal process and the statistical treatment provide an improvement by a factor of 2 in the expected limit in the case of the $\PH\to\PJGy\PGg$ decay, and of 30\% in the case of the $\PZ\to\PJGy\PGg$ decay.

\section{Summary}\label{sec:summary}

A search for rare decays of the \PZ and \PH bosons to a photon and a \PGyPnS meson decaying to a pair of muons is presented.
The data were collected by the CMS experiment in proton-proton collisions at $\sqrt{s}=13\TeV$ during 2016--2018 and correspond to an integrated luminosity of 123\fbinv.
No statistically significant excess has been observed over the standard model expectations.
Upper limits at 95\% confidence level are set on the branching fractions: $0.6 \times 10^{-6}$ and $1.3 \times 10^{-6}$ for the $\PZ\to\PJGy\PGg$ and $\PZ\to\PGyP{2S}\PGg$ decays, respectively, and $2.6 \times 10^{-4}$ and $9.9 \times 10^{-4}$ for the $\PH\to\PJGy\PGg$ and $\PH\to\PGyP{2S}\PGg$ decays, respectively.
The limit for $\PH\to\PJGy\PGg$ translates to an interval constraint on the $\kc/\kg$ ratio of the Higgs boson coupling modifiers of $(-157,\, +199)$. If the standard model value of $\kg=1$ is assumed, the observed interval becomes $(-166,\, +208)$.

\begin{acknowledgments}
  We congratulate our colleagues in the CERN accelerator departments for the excellent performance of the LHC and thank the technical and administrative staffs at CERN and at other CMS institutes for their contributions to the success of the CMS effort. In addition, we gratefully acknowledge the computing centers and personnel of the Worldwide LHC Computing Grid and other centers for delivering so effectively the computing infrastructure essential to our analyses. Finally, we acknowledge the enduring support for the construction and operation of the LHC, the CMS detector, and the supporting computing infrastructure provided by the following funding agencies: SC (Armenia), BMBWF and FWF (Austria); FNRS and FWO (Belgium); CNPq, CAPES, FAPERJ, FAPERGS, and FAPESP (Brazil); MES and BNSF (Bulgaria); CERN; CAS, MoST, and NSFC (China); MINCIENCIAS (Colombia); MSES and CSF (Croatia); RIF (Cyprus); SENESCYT (Ecuador); ERC PRG, RVTT3 and MoER TK202 (Estonia); Academy of Finland, MEC, and HIP (Finland); CEA and CNRS/IN2P3 (France); SRNSF (Georgia); BMBF, DFG, and HGF (Germany); GSRI (Greece); NKFIH (Hungary); DAE and DST (India); IPM (Iran); SFI (Ireland); INFN (Italy); MSIP and NRF (Republic of Korea); MES (Latvia); LMTLT (Lithuania); MOE and UM (Malaysia); BUAP, CINVESTAV, CONACYT, LNS, SEP, and UASLP-FAI (Mexico); MOS (Montenegro); MBIE (New Zealand); PAEC (Pakistan); MES and NSC (Poland); FCT (Portugal); MESTD (Serbia); MCIN/AEI and PCTI (Spain); MOSTR (Sri Lanka); Swiss Funding Agencies (Switzerland); MST (Taipei); MHESI and NSTDA (Thailand); TUBITAK and TENMAK (Turkey); NASU (Ukraine); STFC (United Kingdom); DOE and NSF (USA).

  \hyphenation{Rachada-pisek} Individuals have received support from the Marie-Curie program and the European Research Council and Horizon 2020 Grant, contract Nos.\ 675440, 724704, 752730, 758316, 765710, 824093, 101115353, 101002207, and COST Action CA16108 (European Union); the Leventis Foundation; the Alfred P.\ Sloan Foundation; the Alexander von Humboldt Foundation; the Science Committee, project no. 22rl-037 (Armenia); the Belgian Federal Science Policy Office; the Fonds pour la Formation \`a la Recherche dans l'Industrie et dans l'Agriculture (FRIA-Belgium); the F.R.S.-FNRS and FWO (Belgium) under the ``Excellence of Science -- EOS" -- be.h project n.\ 30820817; the Beijing Municipal Science \& Technology Commission, No. Z191100007219010 and Fundamental Research Funds for the Central Universities (China); the Ministry of Education, Youth and Sports (MEYS) of the Czech Republic; the Shota Rustaveli National Science Foundation, grant FR-22-985 (Georgia); the Deutsche Forschungsgemeinschaft (DFG), among others, under Germany's Excellence Strategy -- EXC 2121 ``Quantum Universe" -- 390833306, and under project number 400140256 - GRK2497; the Hellenic Foundation for Research and Innovation (HFRI), Project Number 2288 (Greece); the Hungarian Academy of Sciences, the New National Excellence Program - \'UNKP, the NKFIH research grants K 131991, K 133046, K 138136, K 143460, K 143477, K 146913, K 146914, K 147048, 2020-2.2.1-ED-2021-00181, and TKP2021-NKTA-64 (Hungary); the Council of Science and Industrial Research, India; ICSC -- National Research Center for High Performance Computing, Big Data and Quantum Computing and FAIR -- Future Artificial Intelligence Research, funded by the NextGenerationEU program (Italy); the Latvian Council of Science; the Ministry of Education and Science, project no. 2022/WK/14, and the National Science Center, contracts Opus 2021/41/B/ST2/01369 and 2021/43/B/ST2/01552 (Poland); the Funda\c{c}\~ao para a Ci\^encia e a Tecnologia, grant CEECIND/01334/2018 (Portugal); the National Priorities Research Program by Qatar National Research Fund;  MCIN/AEI/10.13039/501100011033, ERDF ``a way of making Europe", and the Programa Estatal de Fomento de la Investigaci{\'o}n Cient{\'i}fica y T{\'e}cnica de Excelencia Mar\'{\i}a de Maeztu, grant MDM-2017-0765 and Programa Severo Ochoa del Principado de Asturias (Spain); the Chulalongkorn Academic into Its 2nd Century Project Advancement Project, and the National Science, Research and Innovation Fund via the Program Management Unit for Human Resources \& Institutional Development, Research and Innovation, grant B39G670016 (Thailand); the Kavli Foundation; the Nvidia Corporation; the SuperMicro Corporation; the Welch Foundation, contract C-1845; and the Weston Havens Foundation (USA).
\end{acknowledgments}

\bibliography{auto_generated}

\cleardoublepage \appendix\section{The CMS Collaboration \label{app:collab}}\begin{sloppypar}\hyphenpenalty=5000\widowpenalty=500\clubpenalty=5000\input{SMP-22-012-public-authorlist.tex}\end{sloppypar}
%%% END EDITABLE REGION %%%
% skeleton_end
\end{document}

%% file: SMP-22-012-public-authorlist.tex
\cmsinstitute{Yerevan Physics Institute, Yerevan, Armenia}
{\tolerance=6000
A.~Hayrapetyan, A.~Tumasyan\cmsAuthorMark{1}\cmsorcid{0009-0000-0684-6742}
\par}
\cmsinstitute{Institut f\"{u}r Hochenergiephysik, Vienna, Austria}
{\tolerance=6000
W.~Adam\cmsorcid{0000-0001-9099-4341}, J.W.~Andrejkovic, T.~Bergauer\cmsorcid{0000-0002-5786-0293}, S.~Chatterjee\cmsorcid{0000-0003-2660-0349}, K.~Damanakis\cmsorcid{0000-0001-5389-2872}, M.~Dragicevic\cmsorcid{0000-0003-1967-6783}, P.S.~Hussain\cmsorcid{0000-0002-4825-5278}, M.~Jeitler\cmsAuthorMark{2}\cmsorcid{0000-0002-5141-9560}, N.~Krammer\cmsorcid{0000-0002-0548-0985}, A.~Li\cmsorcid{0000-0002-4547-116X}, D.~Liko\cmsorcid{0000-0002-3380-473X}, I.~Mikulec\cmsorcid{0000-0003-0385-2746}, J.~Schieck\cmsAuthorMark{2}\cmsorcid{0000-0002-1058-8093}, R.~Sch\"{o}fbeck\cmsorcid{0000-0002-2332-8784}, D.~Schwarz\cmsorcid{0000-0002-3821-7331}, M.~Sonawane\cmsorcid{0000-0003-0510-7010}, S.~Templ\cmsorcid{0000-0003-3137-5692}, W.~Waltenberger\cmsorcid{0000-0002-6215-7228}, C.-E.~Wulz\cmsAuthorMark{2}\cmsorcid{0000-0001-9226-5812}
% ...
\par}
\vskip\cmsinstskip
\dag:~Deceased\\
$^{1}$Also at Yerevan State University, Yerevan, Armenia\\
$^{2}$Also at TU Wien, Vienna, Austria\\
% ...
